\documentclass[10pt,a4paper]{article}
\usepackage{amsmath}
\usepackage{amsfonts}
\usepackage{amssymb}
\usepackage[dvips]{graphicx}
\usepackage{bm}

\begin{document}

\title{\textbf{Generation of a bubble universe using a\\ negative energy bath}}
\author{\textsc{Dong-il Hwang}$^{a}$\footnote{enotsae@gmail.com}\;\; and \textsc{Dong-han Yeom}$^{a,b,c}$\footnote{innocent.yeom@gmail.com}\\
\textit{$^{a}$\small{Department of Physics, KAIST, Daejeon 305-701, Republic of Korea}}\\
\textit{$^{b}$\small{Center for Quantum Spacetime, Sogang University, Seoul 121-742, Republic of Korea}}\\
\textit{$^{c}$\small{Research Institute for Basic Science, Sogang University, Seoul 121-742, Republic of Korea}}}
\maketitle

\begin{abstract}
This paper suggests a model for a bubble universe using buildable false vacuum bubbles.
We study the causal structures of collapsing false vacuum bubbles using double-null simulations.
False vacuum bubbles violate the null energy condition and emit negative energy along the outgoing direction through semi-classical effects.
If there are a few collapsing false vacuum bubbles and they emit negative energy to a certain region, then the region can be approximated by a negative energy bath, which means the region is homogeneously filled by negative energy.
If a false vacuum bubble is generated in the negative energy bath and the tension of the bubble effectively becomes negative in the bath, then the bubble can expand and form an inflating bubble universe.
This scenario uses a set of assumptions different from those in previous studies because it does not require tunneling to unbuildable bubbles.
\end{abstract}

\newpage

\tableofcontents

\newpage

\section{Introduction}

The information loss problem \cite{Hawking:1976ra} is related to the issue of whether or not the quantum theory of gravity is unitary. It is unclear how we can reconstruct the original information of a black hole after the black hole has totally evaporated. If a theory is not unitary, we cannot expect a fundamental predictability from the theory. Therefore, many researchers have tried to explain how a black hole can conserve information during the evaporation \cite{Susskind:1993if}\cite{Ashtekar:2005cj}\cite{Hawking:2005kf}.

An even worse situation occurs if there is a wormhole or a bubble universe that includes a second asymptotic region. If a wormhole metric is given, it will violate unitarity \cite{Hawking:1987mz}; and if there is a sufficient amount of exotic matter, we can maintain the wormhole for quite long time \cite{Morris:1988cz}. However, it is unclear how to build such an initial condition. In many cases, the initial condition already violates unitarity so that a wormhole is not related to the information loss problem.

There have been some discussions suggesting that we can derive a bubble universe from a false vacuum bubble in a true vacuum background \cite{Sato:1981bf}. According to the thin shell approximation, we can obtain a bubble universe solution that inflates inside of a Schwarzschild wormhole \cite{Blau:1986cw}\cite{Aguirre:2005xs}\cite{Alberghi:1999kd}; however, if we assume the null energy condition and global hyperbolicity, whenever we obtain such inflating bubble solutions, the initial conditions suffer from the past incompleteness \cite{Farhi:1986ty}. Therefore, unless we violate the null energy condition, we cannot derive a bubble universe. Of course, if we assume arbitrary exotic matter by hand, we can induce a bubble universe, but it is not justified unless the matter is allowed by a fundamental theory, e.g., string theory. A better idea to overcome the null energy condition was to assume a tunneling, the so-called Farhi-Guth-Guven tunneling \cite{Farhi:1989yr}. The probability of such tunneling was calculated though it was extremely low. However, if it is possible in principle, it may imply that the nature allows a violation of unitarity.

On the other hand, when the geometry of the background is an anti de Sitter space, according to the AdS/CFT correspondence \cite{Maldacena:1997re} (and if we trust the correspondence), the time evolution in the anti de Sitter background should be unitary \cite{Freivogel:2005qh} (this idea may be extended to a de Sitter background \cite{dscft}). Then, we can choose between two possibilities:
\begin{enumerate}
\item The generation of a bubble universe does not necessarily imply a violation of unitarity,
\item Farhi-Guth-Guven tunneling is impossible in such a situation.
\end{enumerate}

For the first possibility, we can guess that although a Schwarzschild wormhole (or black hole) separates two universes and information passes from the first to the second asymptotic region, Hawking radiation from the black hole can emit information to the first asymptotic region. This should imply a copy of information on a space-like hypersurface. Although such a copy is improbable, if no observer can see the copy of information, we can relax; this is known as the black hole complementarity principle \cite{Susskind:1993if}. Although local physics seems to be violated, no one can see the violation, and then, it is ``a perfect crime.''

However, according to our previous papers, we strongly suspect that black hole complementarity is not a correct conjecture \cite{Yeom:2008qw}\cite{Yeom1}. If our opinion is correct, then the reasonable next choice is to think that such exotic tunneling should not be allowed in the anti de Sitter background. Apart from our opinion, the second possibility is also the consensus between people who are involved in holography. Only classically buildable bubbles that do not contain inflating regions are allowed by unitary processes \cite{Freivogel:2005qh}.

In this controversial context, we will argue that a bubble universe solution is possible without assuming unbuildable bubbles. \textit{We will only assume buildable bubbles}, where the term `buildable' means that the bubble does not contain an inflating region. However, we can violate the null energy condition through Hawking radiation of the de Sitter space \cite{Gibbons:1977mu}; a false vacuum bubble will emit Hawking radiation and will violate the null energy condition \cite{Birrell:1982ix}\cite{Takagi:1989re}. Then an outside observer of the bubble can obtain a small region that can be approximated by a negative energy bath, which means the region is homogeneously filled by negative energy. If we further assume that there is a second tunneling of a false vacuum bubble in the negative energy bath, then the region allows expanding and inflating bubble solutions via a violation of the null energy condition, even though the bubble is classically buildable.

In this paper, we will confirm this idea using numerical simulations. We followed the double-null formalism which was developed by previous researchers \cite{Yeom1}\cite{doublenull}\cite{Yeom2}. In particular, to include Hawking radiation of a de Sitter space, we assumed $S$-wave approximation using $1+1$ dimensional results \cite{Birrell:1982ix}\cite{Davies:1976ei}. This approximation gave good results for black holes \cite{Yeom1}\cite{doublenull}.

Section~\ref{sec:dyn} summarizes previous results on dynamics of false vacuum bubbles in the thin shell approximation. In Section~\ref{sec:gen_bath}, we develop a model that is beyond the thin shell approximation for a false vacuum bubble and includes semi-classical effects. Then we discuss the generation of a negative energy bath using false vacuum bubbles. In Section~\ref{sec:gen_bubble}, we discuss a generation of a bubble universe using the negative energy bath and in Section~\ref{sec:cau}, we discuss on our assumptions for bubble universes. Section~\ref{sec:dis} summarizes and contrasts the ideas already known and our new contributions. In Appendix~A, we check the consistency and convergence of our numerical calculations; in Appendix~B, we compare our numerical setup to analytic expectations.

\begin{figure}
\begin{center}
\includegraphics[scale=0.45]{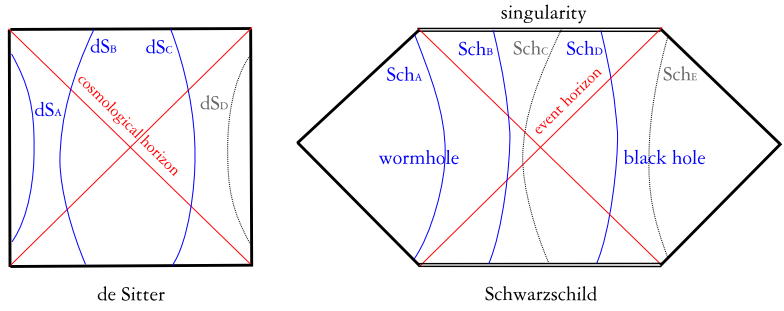}
\caption{\label{fig:thinshell}Solutions of the thin shell approximation (symmetric cases). Since $\beta_{\mathrm{i}}$ is always positive in the $r \rightarrow 0$ limit, $\mathrm{dS}_{\mathrm{D}}$ is disallowed; $\beta_{+}$ is always positive in the $r \rightarrow 0$ limit, $\mathrm{Sch}_{\mathrm{C}}$ is disallowed; $\beta_{+}$ is always negative in the $r \rightarrow \infty$ limit, $\mathrm{Sch}_{\mathrm{E}}$ is disallowed.}
\end{center}
\end{figure}
\begin{figure}
\begin{center}
\includegraphics[scale=0.45]{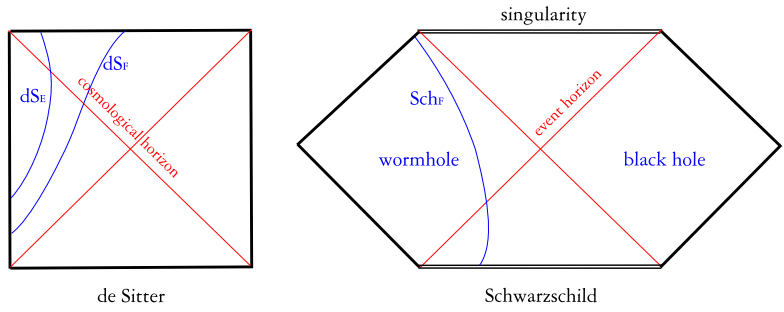}
\caption{\label{fig:thinshell2}Solutions of the thin shell approximation (asymmetric cases).}
\end{center}
\end{figure}
\begin{figure}
\begin{center}
\includegraphics[scale=0.45]{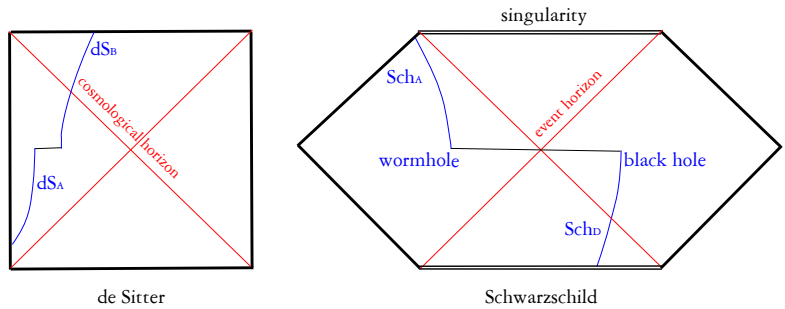}
\caption{\label{fig:FGG}Farhi-Guth-Guven tunneling.}
\end{center}
\end{figure}

\section{\label{sec:dyn}Dynamics of false vacuum bubbles in the thin shell approximation}

If we can apply the thin shell approximation \cite{israel01} with spherical symmetry, the inside can be described by a static de Sitter space and the outside by a static Schwarzschild space: for the inside,
\begin{eqnarray}
ds_{-}^{2} = - f_{-}(R)dT^{2} + \frac{1}{f_{-}(R)}dR^{2} + R^{2}d\Omega^{2},
\end{eqnarray}
and for the outside,
\begin{eqnarray}
ds_{+}^{2} = - f_{+}(R)dT^{2} + \frac{1}{f_{+}(R)}dR^{2} + R^{2}d\Omega^{2},
\end{eqnarray}
where $f_{-}(R) = 1 - R^{2}/l^{2}$ and $f_{+}(R) = 1 - 2M / R$. The remaining part is on the dynamics of the shell at $R=r(\tau)$ and the dynamics is determined by
\begin{eqnarray}
\epsilon_{-}\sqrt{\dot{r}^2 + f_{-}} - \epsilon_{+}\sqrt{\dot{r}^2 + f_{+}} &=& 4\pi r \sigma,\\
\dot{r}^{2} + V_{\mathrm{eff}}(r) &=& 0,
\end{eqnarray}
where
\begin{eqnarray}
V_{\mathrm{eff}}(r) = f_{+} - \frac{(f_{-} - f_{+} - 16 \pi^{2} \sigma^{2} r^{2})^{2}}{64 \pi^{2} \sigma^{2} r^{2}},
\end{eqnarray}
$\tau$ is the time parameter of the shell, $\sigma$ is the tension of the shell and $\epsilon_{\pm}$ are $+1$ if the outward normal to the shell is pointing towards an increasing $r$ and $-1$ if towards a decreasing $r$ \cite{Sato:1981bf}\cite{Blau:1986cw}\cite{Aguirre:2005xs}\cite{Freivogel:2005qh}. In general, the effective potential $V_{\mathrm{eff}}$ is a convex function for a time-like shell \cite{Blau:1986cw}\cite{Aguirre:2005xs}\cite{Freivogel:2005qh}; therefore, it allows a collapsing solution or an expanding solution. There are basically five possibilities: (a) from expanding to collapsing, (b) from collapsing to expanding, (c) from collapsing to collapsing, (d) from expanding to expanding, and (e) a static solution in an unstable equilibrium. Solutions (a) and (b) are symmetric solutions, whereas (c) and (d) are asymmetric solutions. For simplicity, we omit the unstable equilibrium case (e).

To maintain the information of the sign of each of the roots $\epsilon_{\pm}$, we need to compare the extrinsic curvatures for the outside and the inside of the shell \cite{Blau:1986cw}\cite{Aguirre:2005xs}\cite{Freivogel:2005qh}. The extrinsic curvatures are defined as follows:
\begin{eqnarray}
\beta_{-} = \frac{f_{-} - f_{+} + 16 \pi^{2} {\sigma}^{2} r^{2}}{8 \pi r \sigma} = \pm \sqrt{\dot{r}^{2} + f_{-}},
\end{eqnarray}
and
\begin{eqnarray}
\beta_{+} = \frac{f_{-} - f_{+} - 16 \pi^{2} {\sigma}^{2} r^{2}}{8 \pi r \sigma} = \pm \sqrt{\dot{r}^{2} + f_{+}}.
\end{eqnarray}
Now, to satisfy the Einstein equations, the relation
\begin{eqnarray}
\beta_{-} - \beta_{+} = 4 \pi \sigma r
\end{eqnarray}
should hold \cite{Freivogel:2005qh}.

When the null energy condition is satisfied, we can classify all causal structures and some solutions show formation of bubble universes (Figures~\ref{fig:thinshell} and \ref{fig:thinshell2}) \cite{Blau:1986cw}\cite{Aguirre:2005xs}\cite{Freivogel:2005qh}.

First let us classify symmetric solutions. The left diagram of Figure~\ref{fig:thinshell} is for the de Sitter space, and the right diagram is for the Schwarzschild space. For a collapsing case, $\mathrm{dS}_{\mathrm{A}}$ or $\mathrm{dS}_{\mathrm{D}}$ are possible; and $\mathrm{Sch}_{\mathrm{B}}$, $\mathrm{Sch}_{\mathrm{C}}$ or $\mathrm{Sch}_{\mathrm{D}}$ are possible. For an expanding case, $\mathrm{dS}_{\mathrm{B}}$ or $\mathrm{dS}_{\mathrm{C}}$ are also possible; and $\mathrm{Sch}_{\mathrm{A}}$ or $\mathrm{Sch}_{\mathrm{E}}$ are possible. However, according to the behavior of the extrinsic curvatures in the $r\rightarrow 0$ or $r\rightarrow \infty$ limit, we can remove the solutions of $\mathrm{dS}_{\mathrm{D}}$, $\mathrm{Sch}_{\mathrm{C}}$ and $\mathrm{Sch}_{\mathrm{E}}$. Therefore, there are four possible solutions: $\mathrm{dS}_{\mathrm{A}}-\mathrm{Sch}_{\mathrm{B}}$, $\mathrm{dS}_{\mathrm{A}}-\mathrm{Sch}_{\mathrm{D}}$, $\mathrm{dS}_{\mathrm{B}}-\mathrm{Sch}_{\mathrm{A}}$ and $\mathrm{dS}_{\mathrm{C}}-\mathrm{Sch}_{\mathrm{A}}$. Case $\mathrm{dS}_{\mathrm{A}}-\mathrm{Sch}_{\mathrm{B}}$ is a collapsing bubble solution, where the collapsing shell is inside of a Schwarzschild wormhole.
Case $\mathrm{dS}_{\mathrm{A}}-\mathrm{Sch}_{\mathrm{D}}$ is a collapsing bubble solution, where the collapsing shell induces a Schwarzschild black hole. Case $\mathrm{dS}_{\mathrm{B}}-\mathrm{Sch}_{\mathrm{A}}$ is an expanding bubble solution, where the shell expands inside of a Schwarzschild wormhole and the shell becomes greater than the horizon size of the inside de Sitter space. Case $\mathrm{dS}_{\mathrm{C}}-\mathrm{Sch}_{\mathrm{A}}$ is an expanding bubble solution, where the shell expands inside of a Schwarzschild wormhole and the shell expands outside of the cosmological horizon for the $r=0$ observer.

Second, let us classify asymmetric solutions (Figure~\ref{fig:thinshell2}). The most interesting case is the creation of a bubble universe. In this case, we need to consider the expanding to expanding solution. Here, $\mathrm{dS}_{\mathrm{E}}$, $\mathrm{dS}_{\mathrm{F}}$ and $\mathrm{Sch}_{\mathrm{F}}$ are allowed, thus giving us the case of $\mathrm{dS}_{\mathrm{E}}-\mathrm{Sch}_{\mathrm{F}}$ and $\mathrm{dS}_{\mathrm{F}}-\mathrm{Sch}_{\mathrm{F}}$ as allowed transition solutions. We can interpret these as expanding solutions that begin from a singularity.

Note that the initial condition should be singular for such bubble universes, if we assume general relativity, global hyperbolicity and the null energy condition \cite{Farhi:1986ty}. Hence, if a false vacuum bubble includes inflation in its interior, its initial condition is unbuildable in the general relativistic sense. Therefore, to generate a bubble universe, some previous researchers considered tunneling from a classically buildable bubble to a classically unbuildable bubble (Figure~\ref{fig:FGG}) and the tunneling rate could be calculated \cite{Farhi:1989yr}.

\section{\label{sec:gen_bath}Generation of a negative energy bath using false vacuum bubbles}

In this section, we study the dynamics of collapsing false vacuum bubbles. In the thin shell approximation, the dynamics was already studied in the previous section. However, in our analysis, we try two different points: (1) we go beyond the thin shell approximation and (2) we include the semi-classical effects on the false vacuum bubble.

\subsection{\label{sec:set}Setup}

We describe a Lagrangian with a scalar field $\Phi$ and potential $V(\Phi)$ \cite{hawking}\cite{wald}:
\begin{eqnarray} \label{Lagrangian}
\mathcal{L} = - \Phi_{;a}\Phi_{;b}g^{ab}-2V(\Phi).
\end{eqnarray}
From this Lagrangian we can derive the equations of motion for the scalar field:
\begin{eqnarray} \label{scalar}
\Phi_{;ab}g^{ab}-V^{'}(\Phi) = 0.
\end{eqnarray}
In addition, the energy-momentum tensors become
\begin{eqnarray} \label{energy_momentum}
T_{ab}=\Phi_{;a}\Phi_{;b}-\frac{1}{2}g_{ab}(\Phi_{;c}\Phi_{;d}g^{cd}+2V(\Phi)).
\end{eqnarray}

Now, we describe our numerical setup. We start from the double-null coordinates (our convention is $[u,v,\theta,\varphi]$),
\begin{eqnarray} \label{double_null}
ds^{2} = -\alpha^{2}(u,v) du dv + r^{2}(u,v) d\Omega^{2},
\end{eqnarray}
assuming spherical symmetry. Here $u$ is the ingoing null direction and $v$ is the outgoing null direction.

We define the main functions as follows (we follow the numerical approach of previous authors \cite{Hamade:1995ce}\cite{Yeom1}\cite{doublenull}\cite{Yeom2}.): the metric function $\alpha$, the area function $r$, and the massless scalar field $S \equiv \sqrt{4\pi} \Phi$. We also use some conventions: $d \equiv \alpha_{,v}/\alpha$, $h \equiv \alpha_{,u}/\alpha$, $f \equiv r_{,u}$, $g \equiv r_{,v}$, $W \equiv S_{,u}$, $Z \equiv S_{,v}$.

From this setup, the following components can be calculated:
\begin{eqnarray}
G_{uu}&=&-\frac{2}{r} (f_{,u}-2fh), \\
G_{uv}&=&\frac{1}{2r^{2}} \left( 4 rf_{,v} + \alpha^{2} + 4fg \right), \\
G_{vv}&=&-\frac{2}{r} (g_{,v}-2gd), \\
G_{\theta\theta}&=&-4\frac{r^{2}}{\alpha^{2}} \left(d_{,u}+\frac{f_{,v}}{r}\right),
\end{eqnarray}
\begin{eqnarray}
T_{uu}&=&\frac{1}{4\pi} W^{2}, \\
T_{uv}&=&\frac{\alpha^{2}}{2} V(S), \\
T_{vv}&=&\frac{1}{4\pi} Z^{2}, \\
T_{\theta\theta} &=& \frac{r^{2}}{2\pi\alpha^{2}} WZ - r^{2} V(S),
\end{eqnarray}
where
\begin{eqnarray}
V(S) = V(\Phi) |_{\Phi = S/\sqrt{4\pi}}.
\end{eqnarray}

From the equation of the scalar field, we get the following equation:
\begin{eqnarray} \label{scalar_2}
rZ_{,u}+fZ+gW+ \pi \alpha^{2}rV^{'}(S)=0.
\end{eqnarray}
Note that, $V^{'}(S) = dV(S)/dS$.

We also consider renormalized energy-momentum tensors to include semiclassical effects. Spherical symmetry makes it reasonable to use the $1+1$-dimensional results \cite{Davies:1976ei} divided by $4\pi r^{2}$ \cite{Yeom1}\cite{doublenull}:
\begin{eqnarray}
\langle \hat{T}_{uu} \rangle &=& \frac{P}{4\pi r^{2}}\left(h_{,u}-h^{2}\right) \label{Tq1},
 \\
\langle \hat{T}_{uv} \rangle = \langle \hat{T}_{vu} \rangle &=& -\frac{P}{4\pi r^{2}}d_{,u} \label{Tq2},
 \\
\langle \hat{T}_{vv} \rangle &=& \frac{P}{4\pi r^{2}}\left(d_{,v}-d^{2}\right) \label{Tq3},
\end{eqnarray}
with $P \equiv Nl_{\mathrm{Pl}}^2 / 12\pi$, where $N$ is the number of massless scalar fields and $l_{\mathrm{Pl}}$ is the Planck length.
We use the semi-classical Einstein equation,
\begin{eqnarray}
G_{\mu\nu}=8\pi \left( T_{\mu\nu}+\langle \hat{T}_{\mu\nu} \rangle \right).
\end{eqnarray}

Finally, we summarize our simulation equations:
\begin{enumerate}
\item \emph{Einstein equations:}
\begin{eqnarray}
d_{,u} = h_{,v} &=& \frac{1}{1-\frac{P}{r^{2}}} \left[ \frac{fg}{r^{2}} + \frac{\alpha^2}{4r^{2}} -WZ \right], \label{sing} \\
g_{,v} &=& 2dg - rZ^{2} - \frac{P}{r}(d_{,v}-d^{2}), \label{rvv}\\
g_{,u} = f_{,v} &=& -\frac{fg}{r} - \frac{\alpha^{2}}{4r} + 2\pi\alpha^2 r V(S) - \frac{P}{r}d_{,u}, \label{ruv}\\
f_{,u} &=& 2fh - rW^{2} - \frac{P}{r}(h_{,u}-h^{2}) \label{ruu}.
\end{eqnarray}
\item \emph{Scalar field equations:}
\begin{eqnarray}
Z_{,u} = W_{,v} = - \frac{fZ}{r} - \frac{gW}{r} - \pi \alpha^{2}V^{'}(S).
\end{eqnarray}
\end{enumerate}

\subsection{\label{sec:ini}Initial conditions}

First, we use the following potential:
\begin{eqnarray}
V(S) = \left\{ \begin{array}{ll}
0 & S \leq 0, \\
\frac{\Lambda}{2} \left[ 1 - \cos(\pi\frac{S}{\omega}) \right] & 0 < S \leq \omega, \\
\Lambda & \omega < S.
\end{array} \right.
\end{eqnarray}

We prepare a false vacuum bubble along the initial outgoing surface. We need the initial conditions for each function on the initial $u=u_{\mathrm{i}}=0$ and $v=v_{\mathrm{i}}=0$ surfaces. There are two kinds of information: geometry ($\alpha, r, g, f, h, d$) and matter ($S, W, Z$).

On the geometry side, we have gauge freedom to choose the initial $r$ function; although all constant $u$ and $v$ lines are null, there remains freedom to choose the distances between the null lines. We choose $r(u,v_{\mathrm{i}})=ur_{u0}+r_{0}$ and $r(u_{\mathrm{i}},v)=vr_{v0}+r_{0}$. Here, we fix $r_{0}=10$. Then, $g(u_{\mathrm{i}},v)=r_{v0}$ and $f(u,v_{\mathrm{i}})=r_{u0}$ are naturally obtained. For convenience, we choose $r_{u0}=-1/2$ and $r_{v0}=1/2$.
In general, the metric function $\alpha(u,v_{\mathrm{i}})$ has a smaller value than that of a pure de Sitter space if we study the inside of a false vacuum bubble (see the discussion in Section~$4.2.3$ in \cite{Yeom2}). In terms of the mass function $m(u,v) = (r/2) (1+4r_{,u}r_{,v}/\alpha^{2}-8 \pi V r^{2}/3)$ \cite{Waugh:1986jh}, $\alpha(u_{\mathrm{i}},v_{\mathrm{i}})=(1-8 \pi V r_{0}^{2}/3)^{-1/2} > 1$ is the choice for a pure de Sitter space, and hence it is reasonable to choose $\alpha(0,0)=1$ to study the inside of a false vacuum bubble.

On the matter side, first we fix $S(u,v_{\mathrm{i}})=S_{0}$. Then $S(u_{\mathrm{i}},v)$ will be defined by
\begin{eqnarray}
S(u_{\mathrm{i}},v) = \left\{ \begin{array}{ll}
S_{0} & v < v_{\mathrm{shell}},\\
(S_{0} + S_{-}) G(v) - S_{-} & v_{\mathrm{shell}} \leq v < v_{\mathrm{shell}}+\Delta v,\\
- S_{-} & v_{\mathrm{shell}}+\Delta v \leq v,
\end{array} \right.
\end{eqnarray}
where
\begin{eqnarray}
G(v) = 1-\sin^{2} \left[\frac{\pi(v-v_{\mathrm{shell}})}{2\Delta v}\right]
\end{eqnarray}
and where $S_{-}$ is a constant. We fix $S_{0}=\omega$, $v_{\mathrm{shell}}=10$ and $\Delta v=0.5$ for convenience. Then, we can calculate $S(u,v_{\mathrm{i}})$, $W(u,v_{\mathrm{i}})$, $S(u_{\mathrm{i}},v)$ and $Z(u_{\mathrm{i}},v)$ for the initial surfaces. Then, as we fix $S(u_{\mathrm{i}},v)$, from the Einstein equations, we can obtain $\alpha(u_{\mathrm{i}},v)$ from $2gd = rZ^{2}$ (since $r_{,vv}=0$ at the initial surface and asymptotically we can assume that $P=0$). Finally, the other functions can be evolved using equations on $\alpha_{,uv}$, $r_{,uu}$ or $r_{,vv}$ and $S_{,uv}$.

The remaining free parameters are $P$ (the strength of Hawking radiation), $S_{0}$ and $S_{-}$ (the amplitudes of the scalar field) and $\Lambda$ (the vacuum energy of the inside false vacuum region).

We can choose the evolution equation among Equations~(\ref{rvv}), (\ref{ruv}) or (\ref{ruu}) to obtain the function $r$. In this paper, we used Equation~(\ref{ruv}). Then the remaining equations become constraint equations. We obtained $r$ by Equation~(\ref{rvv}) and (\ref{ruu}) and compared it with the original result to check the consistency of the simulations (Appendix~\ref{sec:con}). In this paper, the numerical code was described by the 2nd order Runge-Kutta method \cite{nr}. Its convergence was tested in Appendix~\ref{sec:con}.

\begin{figure}
\begin{center}
\includegraphics[scale=0.4]{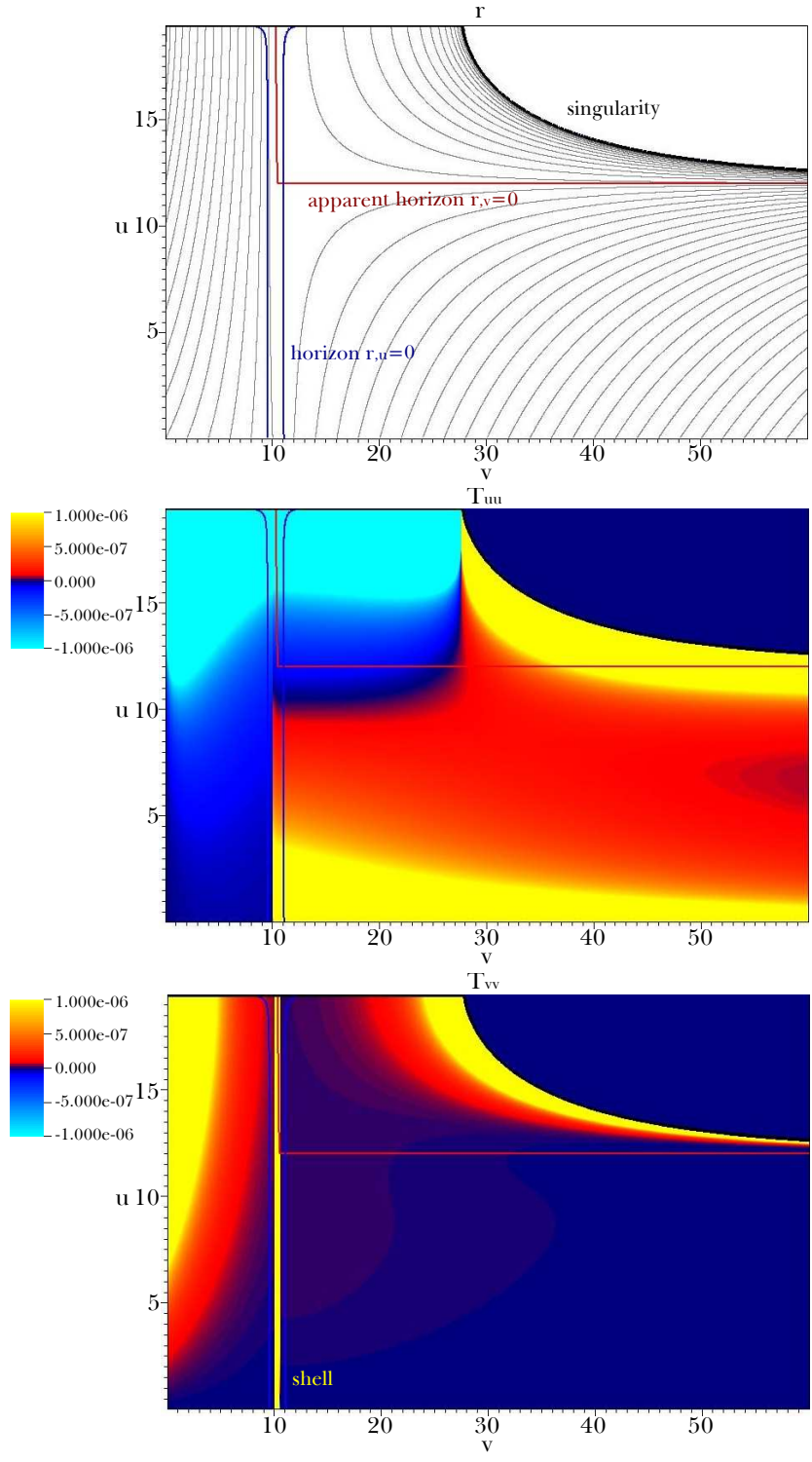}
\caption{\label{fig:case1}$r$, $T_{uu}$, and $T_{vv}$ for $P=0.1$, $S_{0}=0.01$, $S_{-}=0.1$, and $\Lambda = 0.0008$.}
\end{center}
\end{figure}

\begin{figure}
\begin{center}
\includegraphics[scale=0.4]{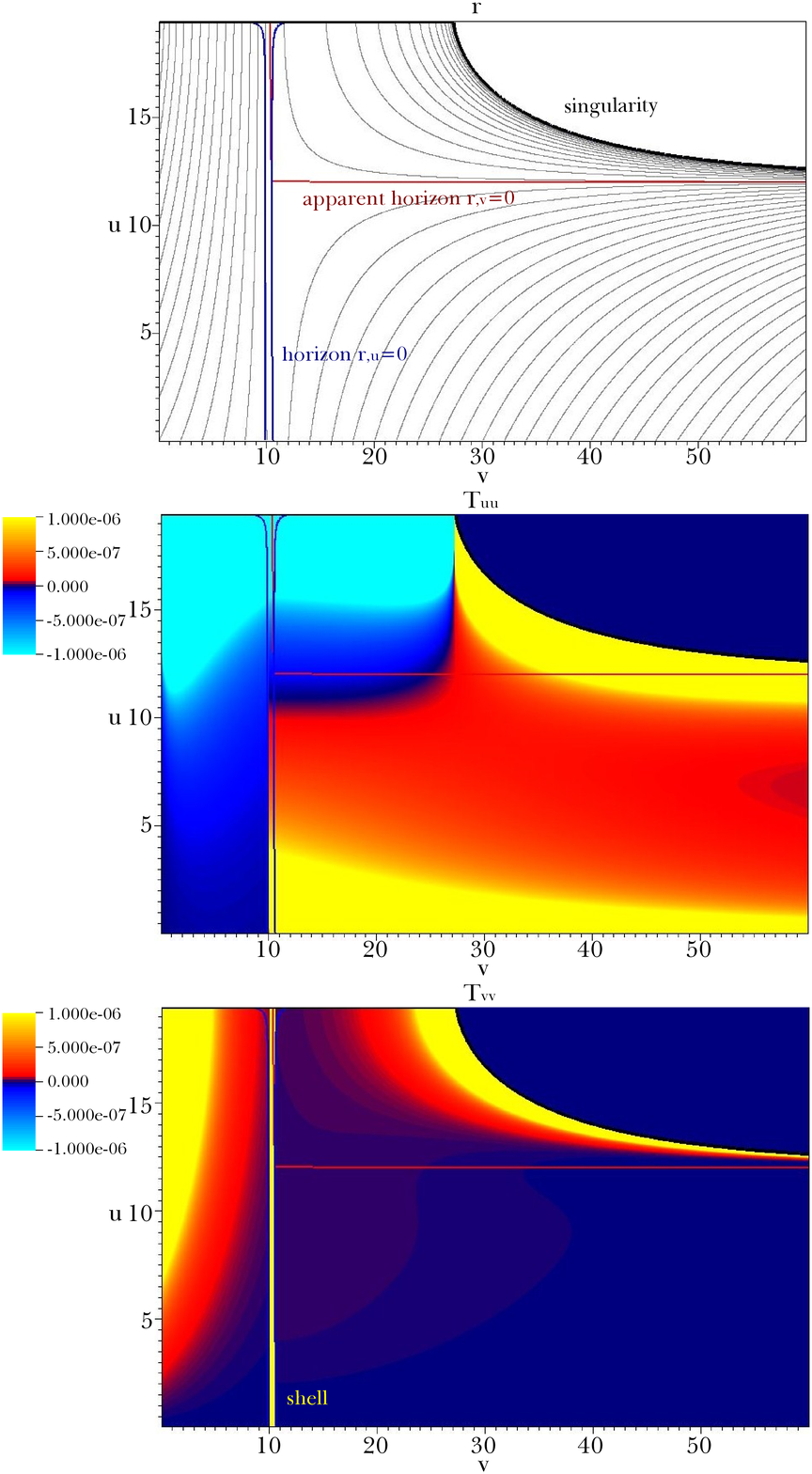}
\caption{\label{fig:case2}$r$, $T_{uu}$, and $T_{vv}$ for $P=0.1$, $S_{0}=0.01$, $S_{-}=0.1$, and $\Lambda = 0.00077$.}
\end{center}
\end{figure}

\begin{figure}
\begin{center}
\includegraphics[scale=0.4]{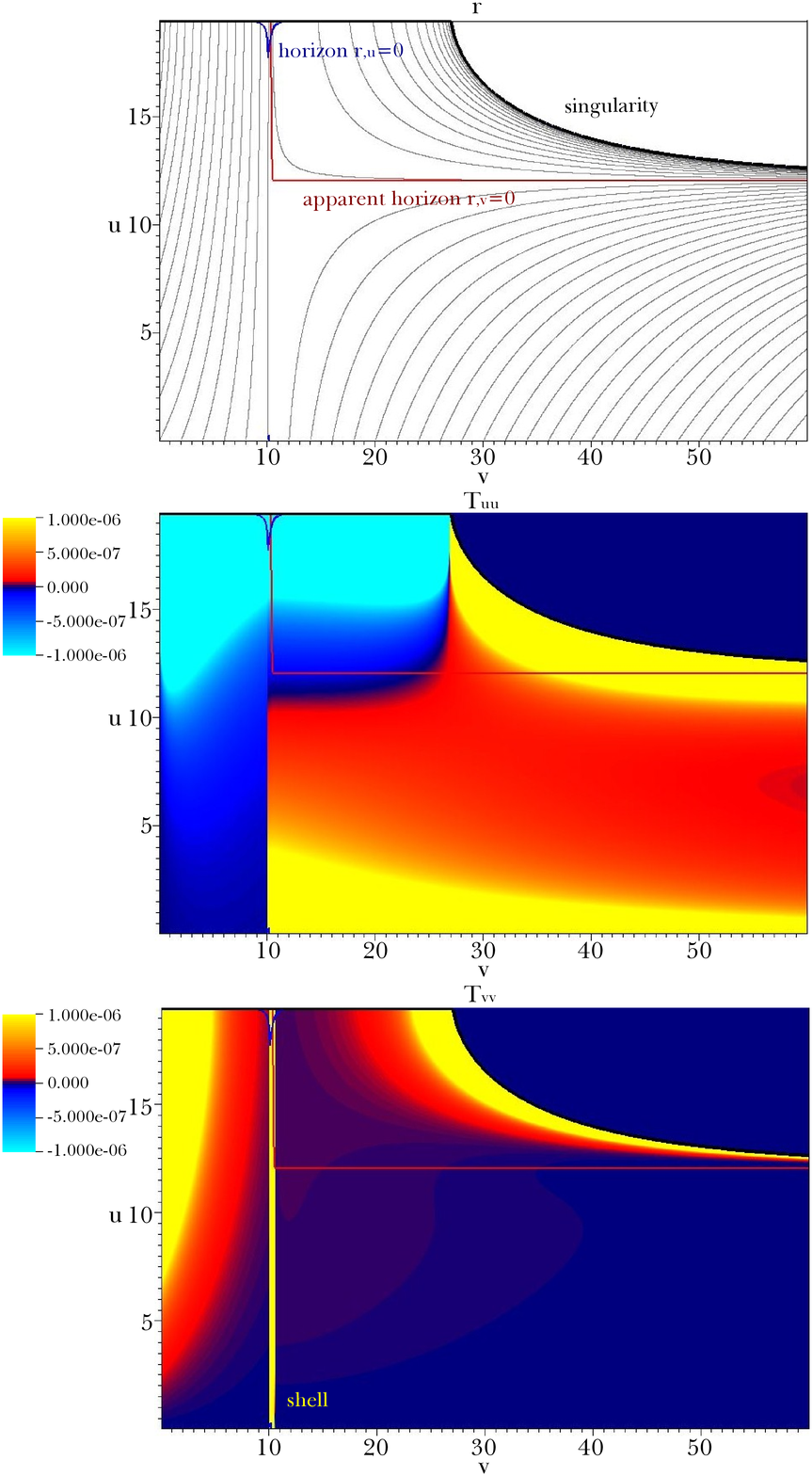}
\caption{\label{fig:case3}$r$, $T_{uu}$, and $T_{vv}$ for $P=0.1$, $S_{0}=0.01$, $S_{-}=0.1$, and $\Lambda = 0.00075$.}
\end{center}
\end{figure}

\begin{figure}
\begin{center}
\includegraphics[scale=0.4]{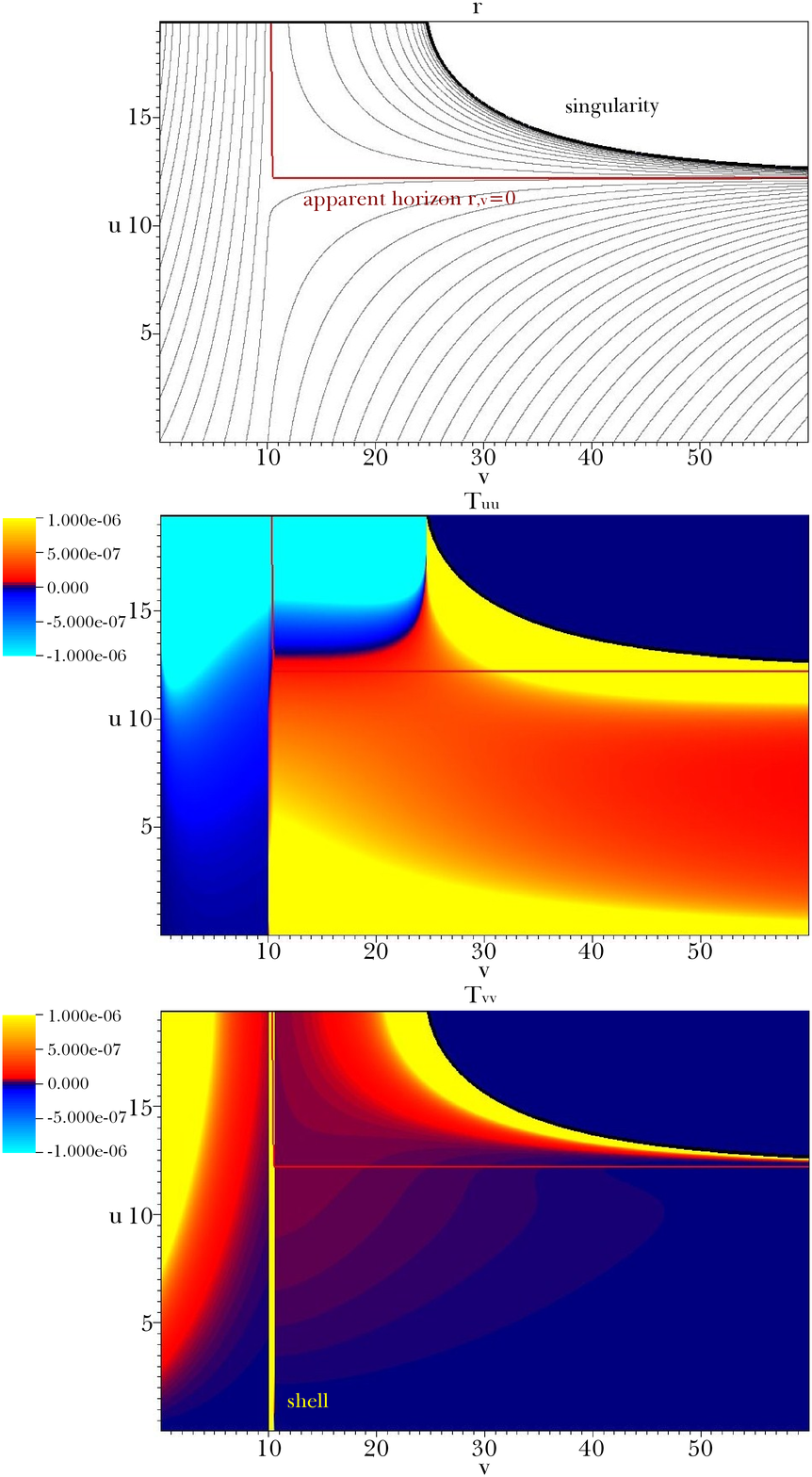}
\caption{\label{fig:case4}$r$, $T_{uu}$, and $T_{vv}$ for $P=0.1$, $S_{0}=0.01$, $S_{-}=0.1$, and $\Lambda = 0.0006$.}
\end{center}
\end{figure}

\subsection{\label{sec:res}Results}

We report the simulation results in Figures~\ref{fig:case1}, \ref{fig:case2}, \ref{fig:case3} and \ref{fig:case4}. The parameters are $P=0.1$, $S_{0}=0.01$, $S_{-}=0.1$ and $\Lambda=0.0008$, $0.00077$, $0.00075$ and $0.0006$.

In Figure~\ref{fig:case1}, we can see two horizons: $r_{,v}=0$ and $r_{,u}=0$. The former is the apparent horizon due to the collapse of the shell and there is a space-like singularity inside of the horizon. The latter is the horizon of the false vacuum bubble. As the shell collapses, the $r_{,u}>0$ region disappears.

\begin{figure}
\begin{center}
\includegraphics[scale=0.6]{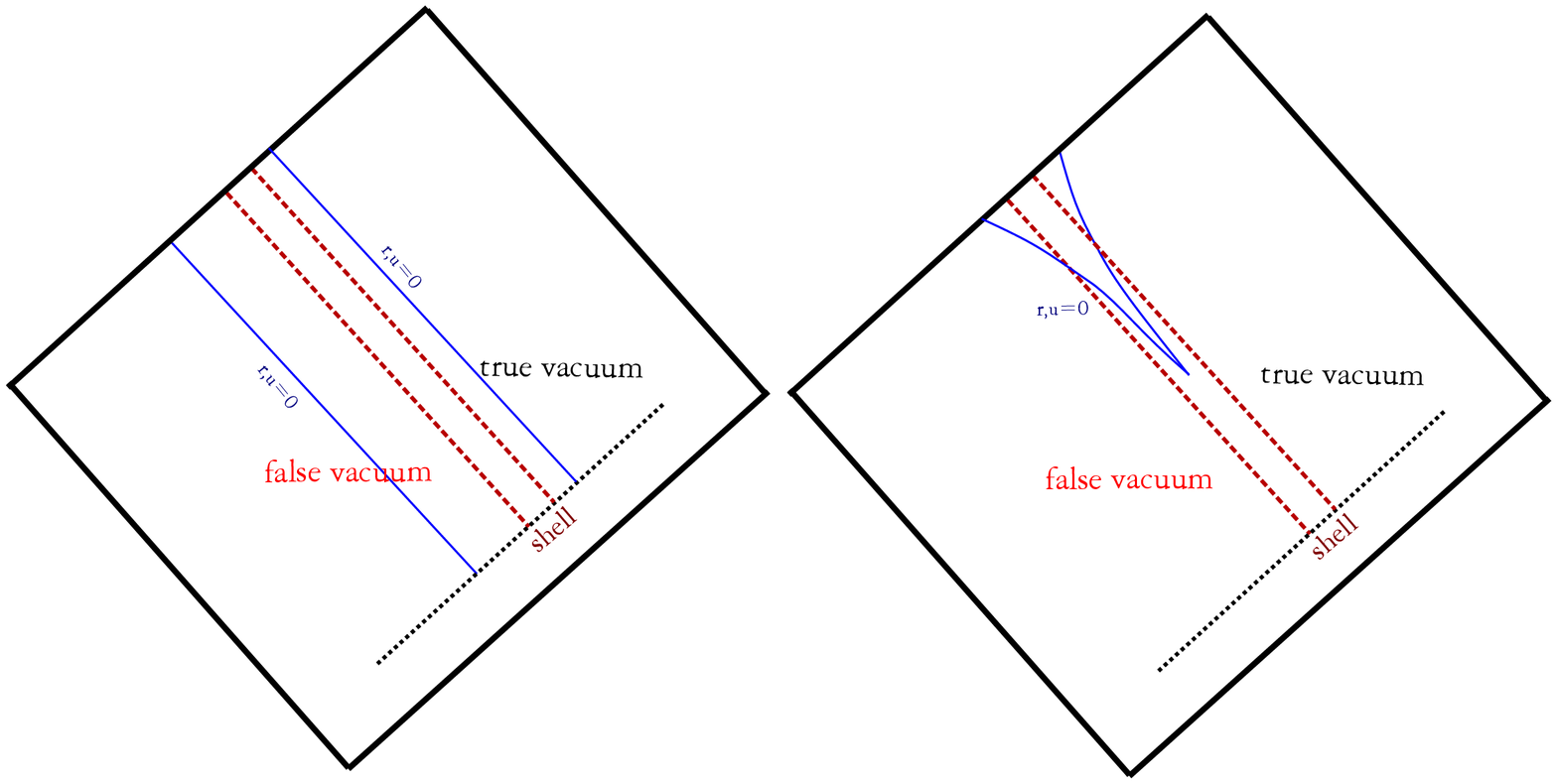}
\caption{\label{fig:transition}As $\Lambda$ decreases or the size of the shell becomes smaller, if there is no violation of the null energy condition, two $r_{,u}=0$ horizons will not meet (left). However, if there is a violation of the null energy condition, two horizons can be combined (right).}
\end{center}
\end{figure}

If the size of the shell is smaller than the horizon size of the false vacuum bubble, we will not see the $r_{,u}=0$ horizon (Figure~\ref{fig:case4}). Between the two limits (Figures~\ref{fig:case1} and \ref{fig:case4}), one would expect to see a smooth transition: Figures~\ref{fig:case2} and \ref{fig:case3} show the proper transition. If there is no violation of the null energy condition, $r_{,u}$ should decrease for all ingoing observers since
\begin{eqnarray}
-\frac{r}{2} G_{uu} = r_{,uu}
\end{eqnarray}
at the $r_{,u}=0$ horizon. Hence, if $G_{uu}>0$, then $r_{,u}$ cannot increase. Then it is difficult to combine the two limits in Figures~\ref{fig:case1} and \ref{fig:case4}. However, in our setup, we introduced a violation of the null energy condition and the violation smoothly connects the two limits (Figure~\ref{fig:transition}). Near the initial surface, we can see a piece of $r_{,u}=0$ horizons in Figure~\ref{fig:case3}. This is also a remainder during the transition, but it is negligible in the whole causal structure.

\begin{figure}
\begin{center}
\includegraphics[scale=0.75]{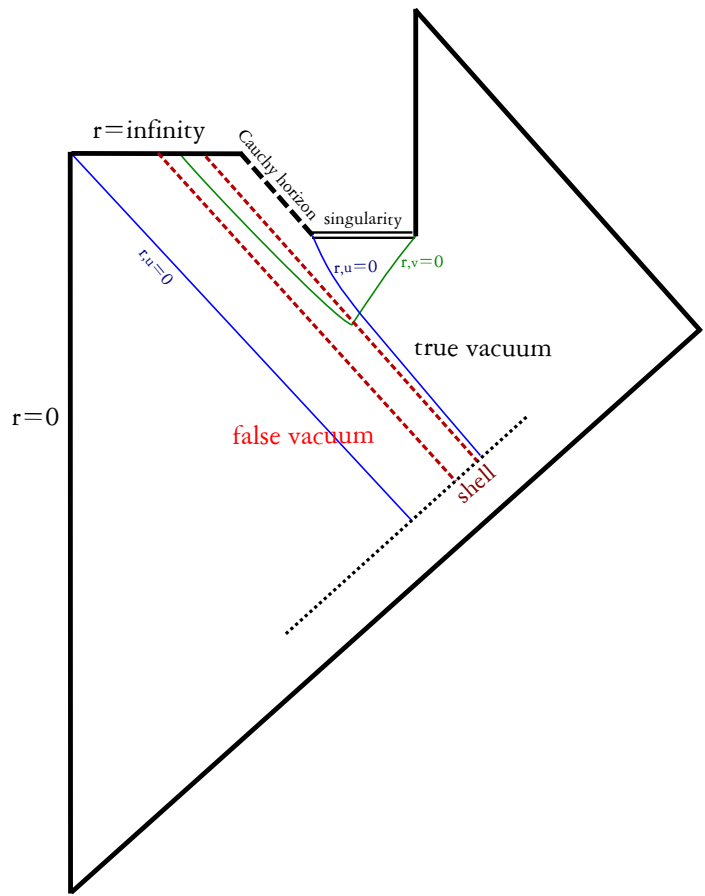}
\caption{\label{fig:ingoing_shell_1}A large unbuildable collapsing shell. Intuitively similar to $\mathrm{dS}_{\mathrm{C}}-\mathrm{Sch}_{\mathrm{A}}$}
\end{center}
\end{figure}

\begin{figure}
\begin{center}
\includegraphics[scale=0.75]{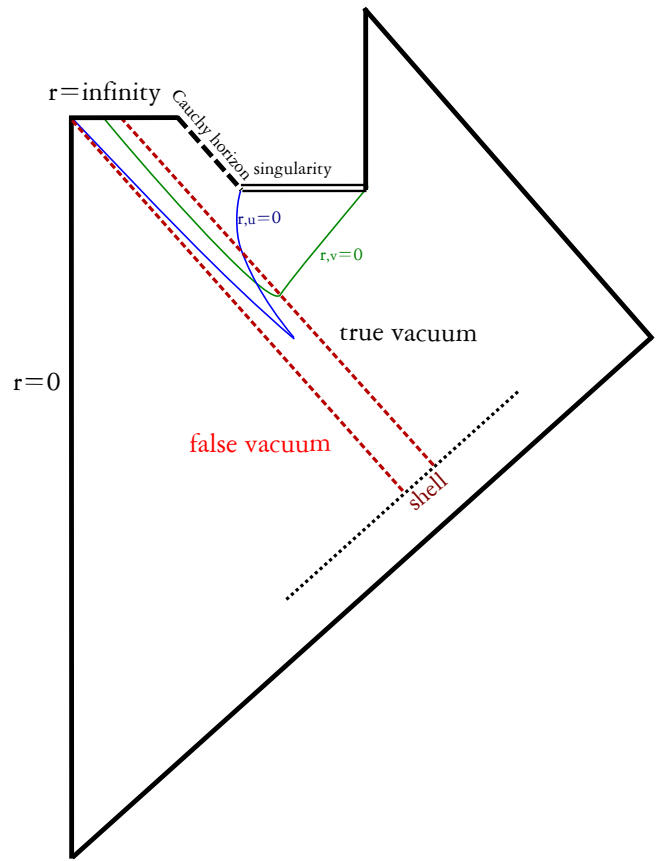}
\caption{\label{fig:ingoing_shell_2}An intermediate collapsing shell.}
\end{center}
\end{figure}

\begin{figure}
\begin{center}
\includegraphics[scale=0.75]{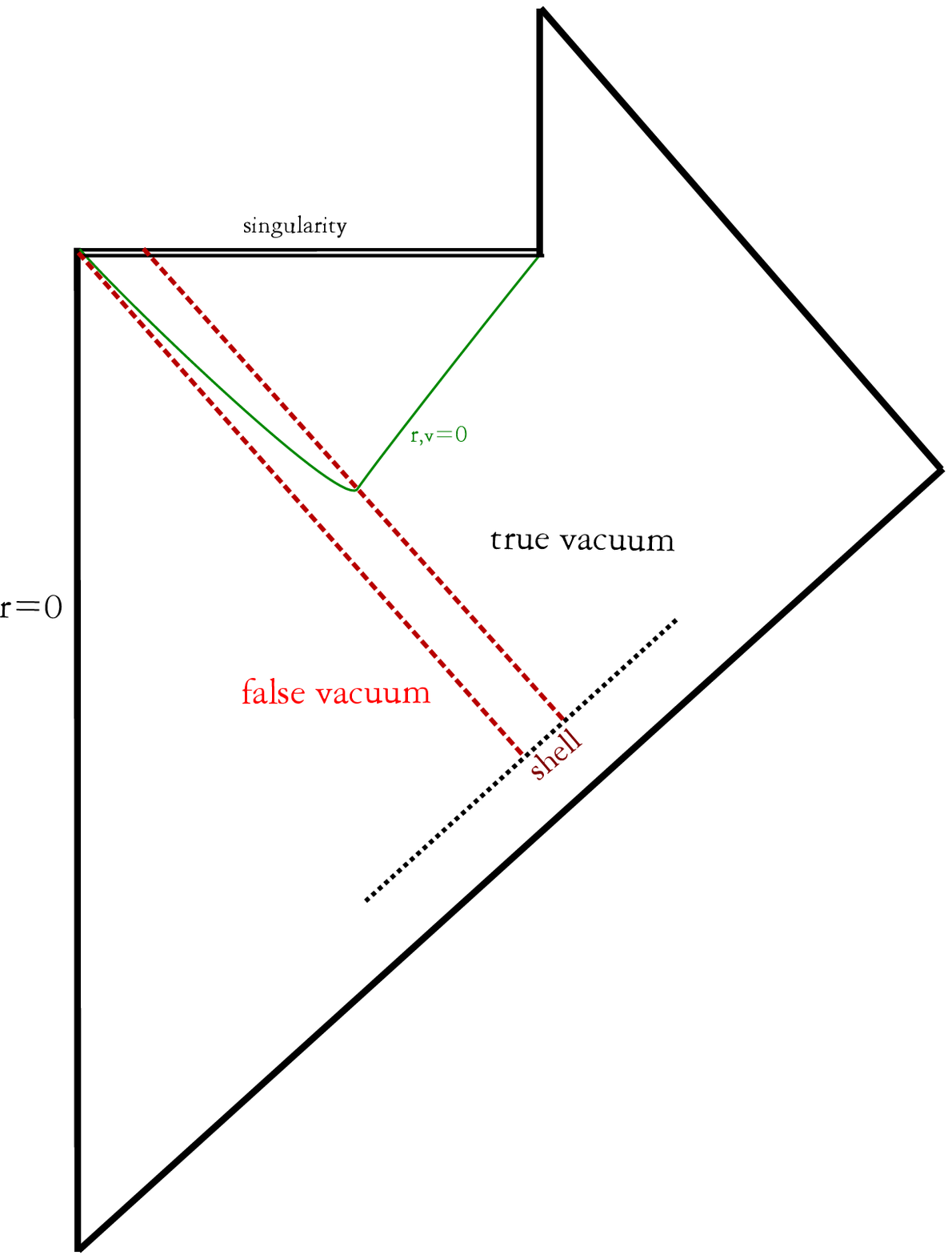}
\caption{\label{fig:ingoing_shell_3}A small buildable collapsing shell. Intuitively similar to $\mathrm{dS}_{\mathrm{A}}-\mathrm{Sch}_{\mathrm{D}}$.}
\end{center}
\end{figure}

Figures~\ref{fig:ingoing_shell_1}, \ref{fig:ingoing_shell_2} and \ref{fig:ingoing_shell_3} are interpretations of the results.

For Figure~\ref{fig:ingoing_shell_1}, the shell is already larger than the horizon size of the false vacuum bubble and it will eventually form the future infinity. As the shell collapses, it will form a black hole and the outside observer will see just a true vacuum region. Then to join the inside and outside, an $r_{,u}>0$ region appears and disappears. There should be a region where the singularity and the future infinity meet; in that region, some complicated things happen, so there should be a Cauchy horizon to connect the singularity and the future infinity. This diagram is intuitively similar to $\mathrm{dS}_{\mathrm{C}}-\mathrm{Sch}_{\mathrm{A}}$.

For Figure~\ref{fig:ingoing_shell_3}, the shell is sufficiently small so that we cannot see the $r_{,u}=0$ horizons. Therefore, this is a buildable solution. This diagram is intuitively similar to $\mathrm{dS}_{\mathrm{A}}-\mathrm{Sch}_{\mathrm{D}}$.

Between the two diagrams (Figures~\ref{fig:ingoing_shell_1} and \ref{fig:ingoing_shell_3}), there can be Figure~\ref{fig:ingoing_shell_2} due to the violation of the null energy condition. The horizon dynamics occurs on the shell and hence it cannot be described by the thin shell approximation. Therefore, this is a new diagram beyond the thin shell approximation and demonstrates the semi-classical effects.

If we see the energy-momentum tensors $T_{uu}$ and $T_{vv}$, we can easily notice that $T_{uu}<0$ and $T_{vv}>0$ inside of the shell. This implies that the false vacuum bubble emits negative energy along the outgoing direction and emits positive energy along the ingoing direction; the situation is opposite for a black hole that emits negative energy along the ingoing direction and emits positive energy along the outgoing direction. As the shell collapses, the outgoing negative energy escapes to the outside of the shell.

\begin{figure}
\begin{center}
\includegraphics[scale=0.27]{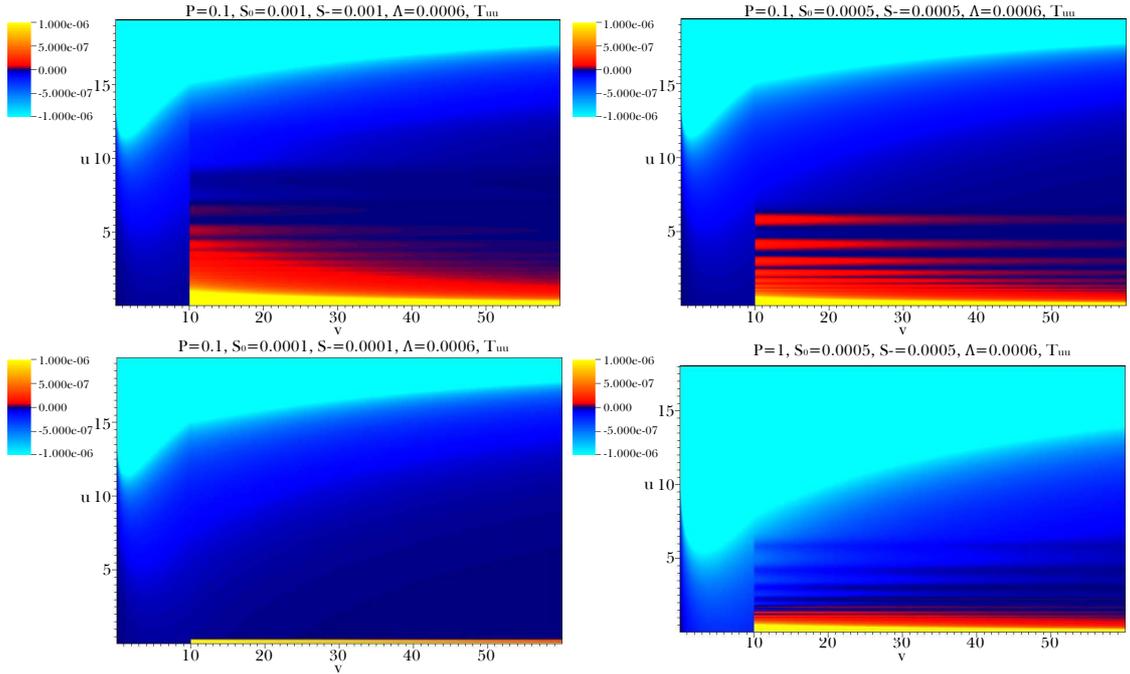}
\caption{\label{fig:various}$T_{uu}$ for various cases. Since the energy of the shell is sufficiently small, we cannot see $r_{,v}=0$ horizons in these figures.}
\end{center}
\end{figure}

One interesting observation is that the collapsing shell initially emits positive energy along the outgoing direction. In the thin shell approximation, there is no such effect and this effect is not an essential one. Even with a thick shell, it is possible to control such effects (Type~$1$ in \cite{Yeom2}). In our setup, we may give too much energy to the shell and hence some part of the energy would be emitted to the outgoing direction. As we choose sufficiently small energy on the shell, the positive energy along the outgoing direction will be controlled. In Figure~\ref{fig:various}, we tested some cases: $P=0.1$, $\Lambda=0.0006$ and varying $S_{0}=S_{-}=0.001$, $0.0005$, and $0.00001$. We also tested the $P=1$, $S_{0}=S_{-}=0.0005$ and $\Lambda=0.0006$ case. Since the energy on the shell $T_{vv} \sim ((S_{0}+S_{-})/\Delta v )^{2}$ decreases, the black hole horizon $r_{,v}=0$ disappeared in the integrated domain. As the energy on the shell decreased, the outgoing positive energy decreased. In addition, as we assume strong Hawking radiation, the outgoing positive energy decreases. Therefore, although there is a positive energy flux along the outgoing direction, it can be controlled and it is not the essential property of a collapsing bubble.

In conclusion, it is clear that a collapsing false vacuum bubble emits negative energy along the outgoing direction and the negative energy can be observed outside of the shell. This is due to the semi-classical effects of the de Sitter space.

\subsection{\label{sec:gen}Generation of a negative energy bath}

To include the semi-classical effect in false vacuum bubbles, we used the $S$-wave approximation of the renormalized energy-momentum tensor of $1+1$ dimensions (metric is $ds^{2}=-\alpha^{2}(u,v)dudv$) \cite{Birrell:1982ix}\cite{Davies:1976ei}:
\begin{eqnarray} \label{semiclassical}
\langle \hat{T}_{uu} \rangle &=& \frac{P}{\alpha^{2}}\left(\alpha \alpha_{,uu} - 2 {\alpha_{,u}}^{2}\right),\\
\langle \hat{T}_{uv} \rangle = \langle \hat{T}_{vu} \rangle &=& -\frac{P}{\alpha^{2}}\left(\alpha\alpha_{,uv}-\alpha_{,u}\alpha_{,v}\right),\\
\langle \hat{T}_{vv} \rangle &=& \frac{P}{\alpha^{2}}\left(\alpha \alpha_{,vv} - 2 {\alpha_{,v}}^{2}\right),
\end{eqnarray}
where $P$ is a constant that is proportional to the number of independent modes of Hawking radiation. For a de Sitter space, $\langle T_{uu} \rangle = \langle T_{vv} \rangle \simeq -P\Lambda$ \cite{Birrell:1982ix}. If we assume the $S$-wave approximation, we will obtain $\langle T_{uu} \rangle = \langle T_{vv} \rangle \simeq -P\Lambda/r^{2}$ in spherically symmetric $3+1$ dimensions. Then, we can see the violation of the null energy condition from the false vacuum bubbles.

This possibility was confirmed by our simulations. Figures~\ref{fig:case1}, \ref{fig:case2}, \ref{fig:case3}, \ref{fig:case4} and \ref{fig:various} as well as their interpretations, Figures~\ref{fig:ingoing_shell_1}, \ref{fig:ingoing_shell_2} and \ref{fig:ingoing_shell_3}, show that negative energy is emitted along the outgoing null direction or a violation of the null energy condition for $T_{uu}$ components. As $\Lambda$ increases, the negative energy increases in Figures~\ref{fig:case1}, \ref{fig:case2}, \ref{fig:case3} and \ref{fig:case4}. In Figure~\ref{fig:various}, as $P$ increases, the amount of negative energy increases.

As $\Lambda$ and $P$ increases, we obtain large violations of the null energy condition. However, $\Lambda$ and $P$ values are limited. First, the size of the shell should be smaller than $l \propto 1/\sqrt{\Lambda}$. Therefore, $\Lambda$ has an upper bound:
\begin{eqnarray}
\Lambda \lesssim \frac{1}{r_{\mathrm{shell}}^{2}}.
\end{eqnarray}
Second, a large $P$ gives the cutoff $\sim \sqrt{P}$ in Equation~(\ref{sing}), which is a boundary of the semi-classical description \cite{Dvali:2007hz}:
\begin{eqnarray}
P \lesssim r_{\mathrm{shell}}^{2}.
\end{eqnarray}
Therefore, the amount of negative energy should be bounded:
\begin{eqnarray}
|\langle T \rangle| \lesssim P \frac{\Lambda}{r^{2}} \lesssim \frac{1}{r^{2}}.
\end{eqnarray}
If we observe a bubble at a sufficiently large distance $r$ from the bubble, then the emitted negative energy will be on the order of $\sim 1/r^{2}$. One note of caution is that even though we apply Dvali's cutoff \cite{Dvali:2007hz}, in fact, we do not have a singularity in the center of the false vacuum region. Therefore, even though we use a sufficiently large $P$, if the center does not have a singularity and the final observing point is larger than $\sqrt{P}$, then the result may still be semi-classical. Thus, it might be possible to obtain a lot of energy from bubbles through a large $P$. 

There is an outgoing positive energy `bounce' which may annihilate such negative energy flow. In our setup, we assumed an ingoing null shell with arbitrary energy on the shell. However, there should be a more proper description using a thin shell with a proper tension: $\mathrm{dS}_{\mathrm{A}}-\mathrm{Sch}_{\mathrm{D}}$. Figures~\ref{fig:case4} and \ref{fig:ingoing_shell_3} correspond to the collapsing part of such a solution. For such a proper limit (and in Figure~\ref{fig:various}), the `bounce' will not annihilate the negative energy.

\begin{figure}
\begin{center}
\includegraphics[scale=0.75]{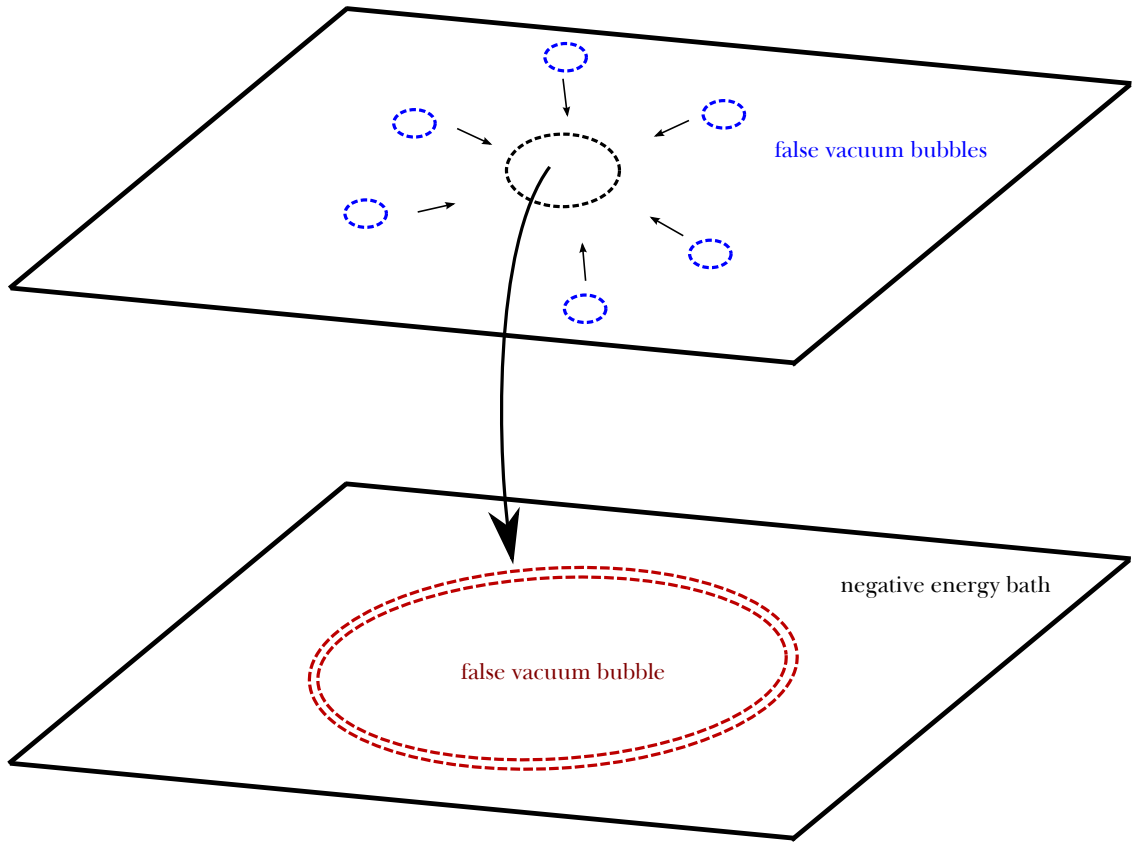}
\caption{\label{fig:bath_form}In the upper figure, buildable false vacuum bubbles (blue dashed circles) generate negative energy along their outgoing directions. Then the energy can be concentrated in a certain region (black dashed circle). If we look at the region locally, it looks like a negative energy bath. In the lower figure, we consider the dynamics of a false vacuum bubble in the negative energy bath.}
\end{center}
\end{figure}

Now, let us imagine the following situation. In a background, many buildable (i.e., does not inflate) false vacuum bubbles are generated and annihilated by collapsing. Then such bubbles will emit negative energy along the outgoing direction. At a distance $r$, the emitted energy will be on the order of $1/r^{2}$. Then, accidently, there can be a certain small region where negative energy of various directions are concentrated (Figure~\ref{fig:bath_form}). In that region, negative energy flows along all directions and appears homogeneous: it seems that there is a bath of negative energy photons. In this region, we can use the energy-momentum tensor
\begin{eqnarray}
T_{\mu\nu} = \mathrm{diag}[-\rho, -p, -p, -p].
\end{eqnarray}
If the amount of $\rho$ and $p$ are not too large so that we can ignore the back-reaction to the background, we can still assume that the background is nearly flat. We call this background a \textit{negative energy bath}.

If a false vacuum bubble is generated and the tension of the bubble is sufficiently small so that the negative energy background can make the tension effectively negative, then the subsequent evolution will be totally different from the original result of pure Einstein gravity.

\section{\label{sec:gen_bubble}Generation of a bubble universe using a negative energy bath}

\subsection{\label{sec:the}The initial setup of a negative energy bath and a bubble}

Let us assume the energy-momentum tensors $T_{\mu \nu}=T^{(1)}_{\mu\nu}+T^{(2)}_{\mu\nu}$:
\begin{eqnarray}
T^{(1)}_{\mu\nu} = \left\{ \begin{array}{ll}
- \Lambda g_{\mu\nu} & \textrm{in false vacuum regions},\\
0 & \textrm{in true vacuum regions},
\end{array} \right.
\end{eqnarray}
with the negative energy bath
\begin{eqnarray}
T^{(2)}_{\mu\nu} = \mathrm{diag}[-\rho, -p, -p, -p].
\end{eqnarray}
Note that, approximately, the negative energy bath contributes to the tension of the shell by $T^{(2)}_{\mu\nu} \epsilon \sim -\rho \epsilon$, where $\epsilon \ll l, M$ (or, $M=0$) is the thickness of the shell. If $|\rho \epsilon| > \sigma$, then the effective tension $\bar{\sigma} \equiv \sigma - \rho \epsilon$ becomes negative.

As long as the shell is in the negative energy bath, such an approximation holds effectively. However, if the shell expands further than the region of the negative energy bath, the tension will be positive again. Therefore, to obtain inflation before the shell reaches the boundary of the negative energy bath, we require that the horizon size of the inside de Sitter space is smaller than the size of the negative energy bath. Of course, it can be chosen as a free parameter \cite{Coleman:1980aw}.

\begin{figure}
\begin{center}
\includegraphics[scale=0.5]{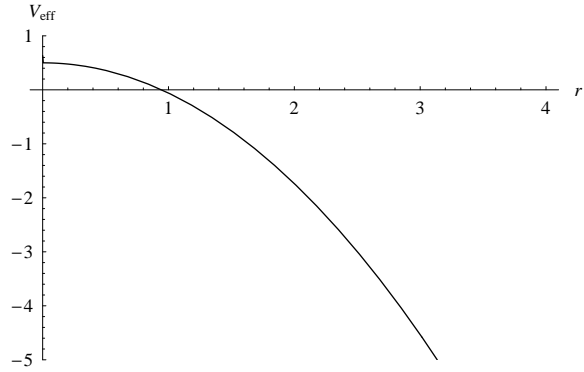}
\caption{\label{fig:l2_sigma_01}An example of $V_{\mathrm{eff}}$, for $l=2$ and $\bar{\sigma}=0.01$. In general, the effective potentials for $M=0$ are qualitatively the same.}
\end{center}
\end{figure}

\begin{figure}
\begin{center}
\includegraphics[scale=0.45]{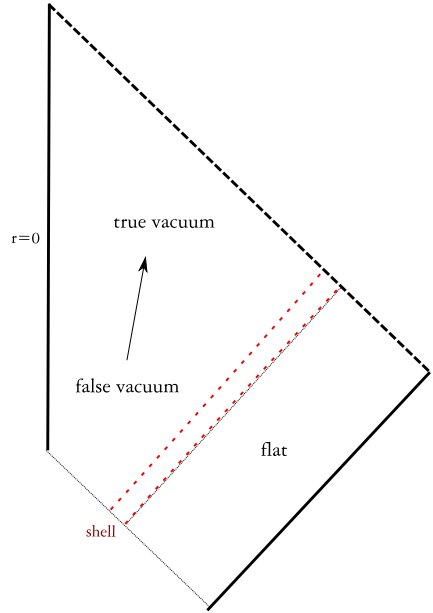}
\caption{\label{fig:unstable}If the tension is positive, a false vacuum bubble in a flat background is unstable and hence buildable.}
\end{center}
\end{figure}
\begin{figure}
\begin{center}
\includegraphics[scale=0.45]{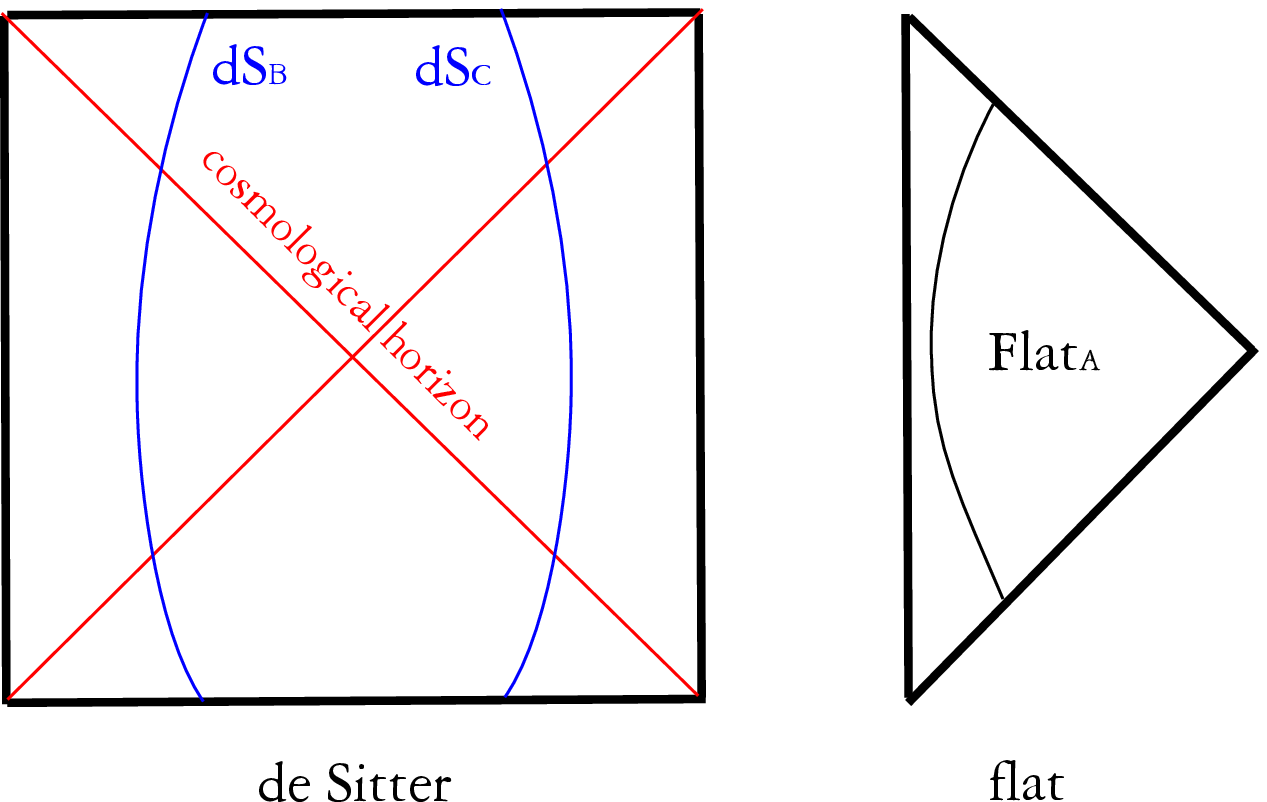}
\caption{\label{fig:thinshell3}Causal structure of a stable expanding bubble. $\mathrm{dS}_{\mathrm{B}}-\mathrm{Flat}_{\mathrm{A}}$ and $\mathrm{dS}_{\mathrm{C}}-\mathrm{Flat}_{\mathrm{A}}$ are both possible.}
\end{center}
\end{figure}
\begin{figure}
\begin{center}
\includegraphics[scale=0.45]{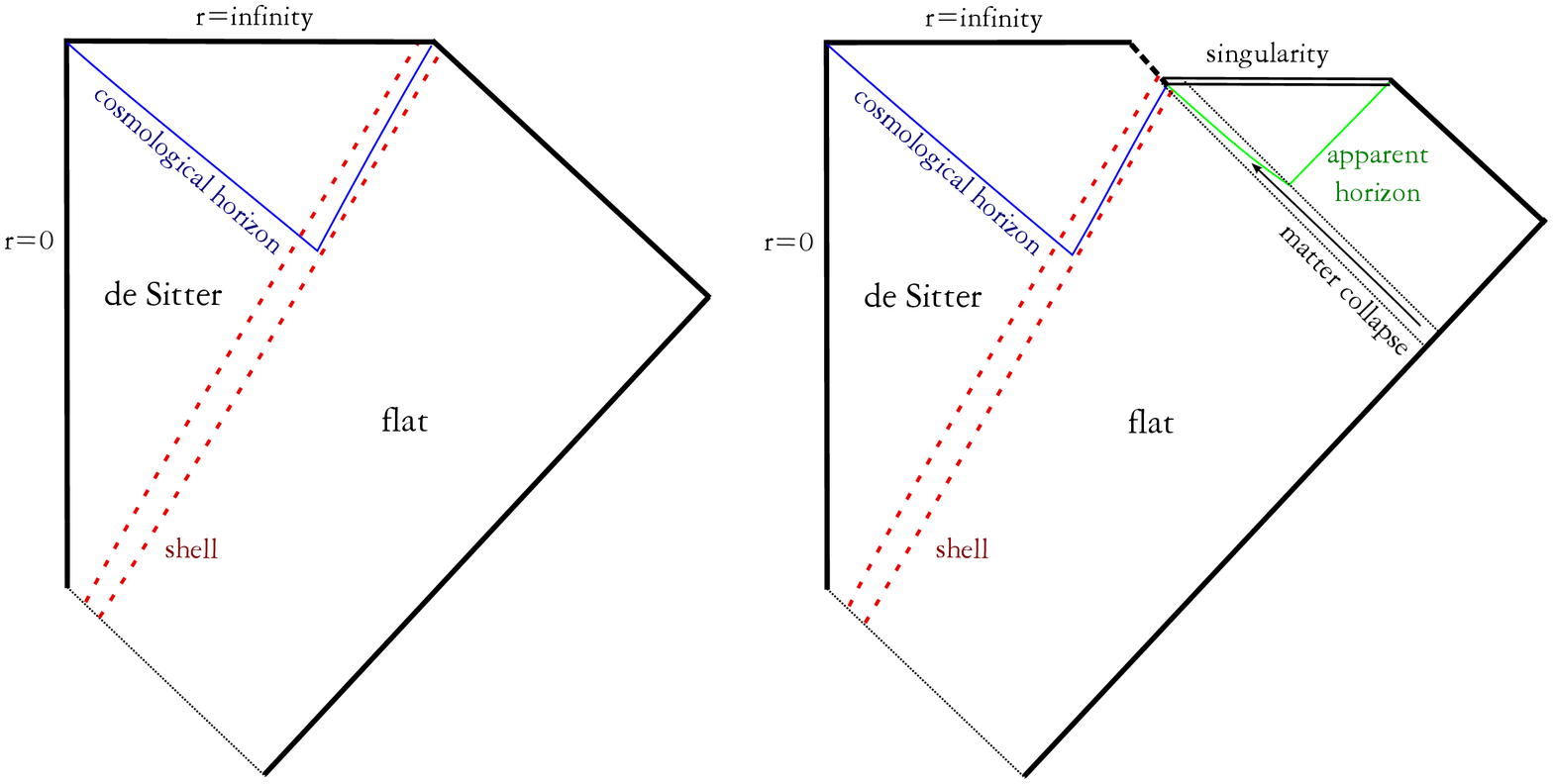}
\caption{\label{fig:stable}If the tension is negative, a false vacuum bubble can expand and inflate. If we send sufficient energy to the bubble, then we can separate the inside inflating space from the outside.}
\end{center}
\end{figure}

\subsection{\label{sec:num}Dynamics of thin shell bubbles and generation of a bubble universe}

Here, we consider bubbles with $M=0$ limit, i.e., bubbles in a nearly flat background. If $\Lambda \sim 1/l^{2} \gg \rho, p$, then the inside and the outside will follow the de Sitter space and the flat space. Now the equation of motion becomes \cite{Blau:1986cw}\cite{Aguirre:2005xs}\cite{Freivogel:2005qh}:
\begin{eqnarray}
\epsilon_{-}\sqrt{\dot{r}^2 + f_{-}} - \epsilon_{+}\sqrt{\dot{r}^2 + f_{+}} &=& 4\pi r \bar{\sigma},\\
\dot{r}^{2} + V_{\mathrm{eff}}(r) &=& 0,
\end{eqnarray}
where
\begin{eqnarray}
V_{\mathrm{eff}}(r) = f_{+} - \frac{(f_{-} - f_{+} - 16 \pi^{2} \bar{\sigma}^{2} r^{2})^{2}}{64 \pi^{2} \bar{\sigma}^{2} r^{2}}
\end{eqnarray}
and $f_{-} = 1-r^{2}/l^{2}$, $f_{+}=1$, $r$ is the radius of the bubble and $\epsilon_{\pm}$ determines the direction of the shell.

To maintain the information of signs of roots $\epsilon_{\pm}$, we use the extrinsic curvatures again:
\begin{eqnarray}
\beta_{-} = \frac{f_{-} - f_{+} + 16 \pi^{2} {\bar{\sigma}}^{2} r^{2}}{8 \pi r \bar{\sigma}} = \pm \sqrt{\dot{r}^{2} + f_{-}},
\end{eqnarray}
and
\begin{eqnarray}
\beta_{+} = \frac{f_{-} - f_{+} - 16 \pi^{2} {\bar{\sigma}}^{2} r^{2}}{8 \pi r \bar{\sigma}} = \pm \sqrt{\dot{r}^{2} + f_{+}}.
\end{eqnarray}

Whether $\bar{\sigma}$ is positive or negative, the effective potential $V_{\mathrm{eff}}(r)$ will have the same form as long as $|\bar{\sigma}|$ and $l$ are the same. In this limit, all bubbles should expand (Figure~\ref{fig:l2_sigma_01}). If $\bar{\sigma}$ is positive, $\beta_{+}$ is always negative and hence it should touch the left boundary of the Minkowski space; or if the bubble is in the right part of the Penrose diagram, the bubble should be unstable (Figure~\ref{fig:unstable}) \cite{Yeom2}. Therefore, such a bubble is buildable and can be generated initially. If $\bar{\sigma}$ is negative, then $\beta_{+}$ is always positive and hence it can touch the right boundary (Figure~\ref{fig:thinshell3}). Therefore, the bubble can inflate and can form a bubble universe (Figure~\ref{fig:stable}). In this case, even though the bubble expands and inflates, the initial state of the bubble is not unbuildable. Rather, in this case, a buildable bubble became an inflating bubble through a violation of the null energy condition.

If the amount of negative energy is sufficiently smaller than the vacuum energy of the inside region, it will not affect the inside of the shell; if it is sufficiently small, it will not affect the outside either; and if the energy-momentum tensor component of the shell is sufficiently smaller than the density of the negative energy, the shell can have a negative tension successively. In fact, it was confirmed that such a situation (negative energy is concentrated on the shell and the tension becomes negative) allows inflation by Type~$3$ in \cite{Yeom2}.

\section{\label{sec:cau}Remarks of caution on bubble universes}

If there are $r_{,u}=0$ horizons, it is not so strange that there is radiation along the ingoing or outgoing directions. However, what will happen if a bubble is slightly smaller than size of the horizon so that one cannot see the horizon? Our calculations confirmed that false vacuum bubbles emit a negative energy flux along the outgoing direction not only as seen in Figures~\ref{fig:ingoing_shell_1} and \ref{fig:ingoing_shell_2} but also in Figure~\ref{fig:ingoing_shell_3} and $\mathrm{dS}_{\mathrm{A}}-\mathrm{Sch}_{\mathrm{D}}$. In the latter cases, the bubbles themselves do not contain inflation and hence they are in principle buildable.

If this is true, we can obtain a negative energy bath using buildable bubbles (Figure~\ref{fig:bath_form}). In the negative energy bath, a buildable bubble can expand and inflate through a violation of the null energy condition (Figures~\ref{fig:thinshell3} and \ref{fig:stable}). If such bubble universes are in principle possible, then it will have theoretical importance for unitarity and holography.

However, such possibilities of creating bubble universes are very rare and accidental. We cautiously remark what are really necessary conditions to obtain such negative energy baths and bubble universes.
\begin{enumerate}
\item Although our false vacuum bubbles are buildable in the sense that there is no inflating region inside of them, it is still unclear whether such buildable false vacuum bubbles can be obtained via quantum tunneling. No one can give a definite answer, but if bounce solutions in the Euclidean signatures are possible, it will be more probable. As we know, there is no known false vacuum bubble bounce solution if the bubble is smaller than the size of the horizon of the background de Sitter space \cite{leeweinberg}\cite{Aguirre:2005xs}.\\
    However, in scalar-tensor gravity, there are some bounce solutions of false vacuum bubbles \cite{nonminimal}, and in some cases, the scalar-tensor gravity will be embedded in string theory \cite{Gasperini:2007zz}. Therefore, although it is rare, string theory may allow such situations.
\item In many cases, we observed not only outgoing negative energy fluxes but also outgoing positive energy fluxes. Of course, in the thin shell limit, we will control such excessive energy to be sufficiently small. However, in realistic and natural situations, the positive energy flux may annihilate negative energy baths.
\item Even though such a negative energy bath is formed, the amount of negative energy will in general be very small. Then, to see the formation of a bubble universe, the tension of the bubble should be extremely small while the inside vacuum energy remains at a constant value. It is not impossible in principle, as long as we tune the potential so that the difference of fields between the two vacuums to be sufficiently small \cite{Coleman:1980aw}. However, it needs a fine-tuning on the potential. If a potential has fundamental constraints, then such small tensions will not be obtained and hence the negative energy bath will be disabled by large tensions.
\item In addition, although such tension is sufficiently small, as the shell expands, if the shell becomes larger than the region of the negative energy bath, the tension will be positive again. Therefore, the effective cosmological constant of the inside de Sitter space should be sufficiently large so that inflation begins before the shell becomes larger than the negative energy bath.\\
    However, if we assume an arbitrarily fine-tuned potential (e.g., landscape \cite{Susskind:2003kw}), such small tension is possible in principle.
\end{enumerate}

Therefore, the scenario of the formation of a bubble universe using a negative energy bath requires favorable chances and fine-tunings of the potential. However, if it is in principle possible, for example in string theory, then it will still have theoretical importance.

\section{\label{sec:dis}Discussion}

We studied the dynamics of collapsing false vacuum bubbles with semi-classical effects beyond the thin shell approximation using double-null simulations. Using this setup, we investigated the possibility of the generation of a negative energy bath using buildable bubbles. We also studied the creation of a bubble universe in a negative energy bath.



There is a novel comment on the causal structure of a collapsing false vacuum bubble. An ingoing observer cannot see $r_{,uu}>0$ at the $r_{,u}=0$ horizon if there is no violation of the null energy condition. If a false vacuum bubble which is slightly larger than the size of the horizon of the internal de Sitter space collapses, the $r_{,u}>0$ region should disappear, and it seems that $r_{,uu}<0$ at the $r_{,u}=0$ horizon. As we violate the null energy condition by renormalized energy-momentum tensors, we can smoothly connect a de Sitter space and a flat space. Therefore, the semi-classical effects regularize and smoothly connect the dynamics of false vacuum bubbles.


Before concluding this paper, we summarize the already known ideas and our new conclusions and suggestions.
The following ideas are already known by previous researchers:
\begin{itemize}
\item[(a)] If tunneling from a buildable bubble to a unbuildable bubble is possible, it will violate unitarity \cite{Farhi:1989yr}; but it is controversial whether or not such tunneling is allowed \cite{Freivogel:2005qh}.
\item[(b)] False vacuum bubbles violate the null energy condition and emit negative energy to an out-going null direction \cite{Gibbons:1977mu}\cite{Birrell:1982ix}\cite{Takagi:1989re}.
\item[(c)] If a violation of the null energy condition is allowed so that the tension of a false vacuum bubble can be negative, then an expanding and inflating bubble is buildable and possible \cite{Blau:1986cw}\cite{Farhi:1986ty}\cite{Freivogel:2005qh}.
\end{itemize}

Now the authors contribute the follow points in this paper:
\begin{itemize}
\item We numerically confirm the emission of the negative energy of false vacuum bubbles, even though the false vacuum bubble is buildable. As we tune initial conditions, the amount of negative energy is sufficiently controllable.
\item The authors connect ideas (b) and (c) above to make a bubble universe, without assuming unbuildable bubbles: if negative energy (emitted by buildable false vacuum bubbles) is concentrated in a region (idea (b)), then it will be possible to see an expanding and inflating bubble in the region (idea (c)).
\end{itemize}

First of all, we reduced one condition for a bubble universe to other simple conditions. Previously, researchers thought that to obtain a bubble universe from buildable bubbles, we needed Farhi-Guth-Guven tunneling. Now we argue that Farhi-Guth-Guven tunneling is not a necessary condition and even though we have buildable bubbles, they violate the null energy condition and it will be possible to form a bubble universe. The other necessary conditions of our scenario are some fine-tunings of the potential. Although it is unnatural, if string theory allows such fine-tunings, this can be a good starting point for interesting discussions on information and unitarity.


\section*{Acknowledgment}
The authors would like to thank Ewan Stewart, Jakob Hansen and Wonwoo Lee for discussions and encouragement.
This work was supported by Korea Research Foundation grants (KRF-313-2007-C00164, KRF-341-2007-C00010) funded by the Korean government (MOEHRD) and BK21. Also, this work was supported by the National Research Foundation of Korea(NRF) grant funded by the Korea government(MEST) through the Center for Quantum Spacetime(CQUeST) of Sogang University with grant number 2005-0049409.

\appendix

\section{\label{sec:con}Consistency and convergence checks}

In this appendix, we discuss convergence and consistency tests for our simulations.

\begin{figure}
\begin{center}
\includegraphics[scale=0.9]{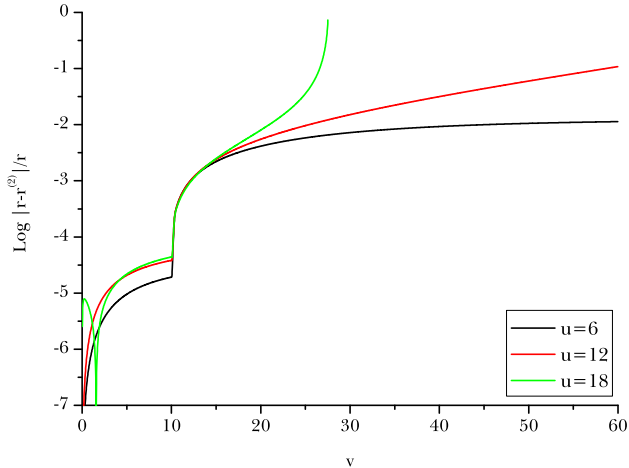}
\caption{\label{fig:consistency}Comparison between two different integration schemes $r$ and $r^{(2)}$, where the former is integrated by Equation~(\ref{ruv}) while the latter is integrated by Equation~(\ref{rvv}) and Equation~(\ref{ruu}). Note that two schemes are equivalent except near the singularity or deep inside of the black hole. The errors are less than 1\% for almost all integrated domains.}
\end{center}
\end{figure}

We mainly obtain the function $r$ by integrating Equation~(\ref{ruv}). However, we can also obtain the function $r$ by integrating Equation~(\ref{rvv}) or (\ref{ruu}). To check the consistency, we used another scheme: obtain $g$ from Equation~(\ref{rvv}) and obtain $f$ from Equation~(\ref{ruu}). We call this result $r^{(2)}$. We compared $r$ and $r^{(2)}$ to check consistency for $u=6$, $12$ and $18$ slices. We used parameters $P=0.1$, $S_{0}=0.01$, $S_{-}=0.1$ and $\Lambda=0.00075$.  Figure~\ref{fig:consistency} shows that they coincide well and the error is less than a few percents except near the singularity or deep inside of the black hole.

\begin{figure}
\begin{center}
\includegraphics[scale=0.9]{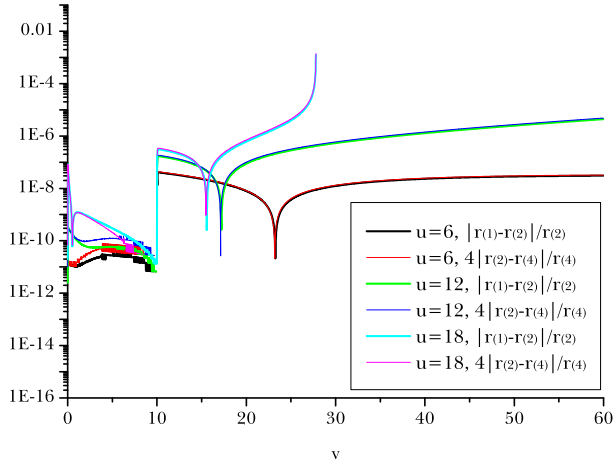}
\caption{\label{fig:convergence}Convergence test. We can see the second order convergence. Errors are less than 1\% for almost all integrated domains.}
\end{center}
\end{figure}

Also, we check the convergence by comparing $1 \times 1$, $2 \times 2$ finer and $4 \times 4$ finer simulations for $u=6$, $12$ and $18$ slices. We used parameters $P=0.1$, $S_{0}=0.01$, $S_{-}=0.1$ and $\Lambda=0.00075$. Figure~\ref{fig:convergence} shows that they converge to the second order and the error is less than 1 \% for almost all integrated domains.

\section{\label{sec:ana}Analytic test for numerical setup}

To trust whether our numerical assumption for the renormalized energy-momentum tensor \cite{Birrell:1982ix}\cite{Davies:1976ei} gives correct physics, a comparison between the analytic and numerical results will be a good test. We know the exact $2$-dimensional results (up to $1$-loop order) and hence Equations~(\ref{Tq1}), (\ref{Tq2}), and (\ref{Tq3}) are the most intuitive and reasonable approximation to the $4$-dimensional cases. For black hole cases, there are studies confirm that this approximation gives correct thermodynamics \cite{doublenull}.

Here, we applied this method to a false vacuum bubble. However, there are some obstacles to comparing the analytic results and the numerical results in the false vacuum background. The known semi-classical results are temperature, entropy, the renormalized energy-momentum tensor components, etc., for the \textit{static limit} or for the de Sitter space. The relevant quantity is to compare the renormalized energy-momentum tensor components of the static limit to our numerical results. Note that, in our setup, we prepare a classical background without Hawking radiation and then we turn on a field combination or Hawking radiation on the background. (Of course, the Einstein equations and field equations are satisfied.) Therefore, energy-momentum tensor components cannot be static and they are essentially time dependent. Moreover, the energy-momentum tensors will give back-reactions to the metric components and these back-reactions will make the comparison difficult.

In this appendix, we compare the energy-momentum tensor components $\langle\hat{T}^{\mathrm{H}}_{uu}\rangle$ as we vary $\Lambda$. We have to determine a space-time position where we make comparison. This is not so well-defined, but one reasonable position is to see the maximum value of $\langle\hat{T}^{\mathrm{H}}_{uu}\rangle$ along an almost asymptotic out-going null surface ($u \simeq 0.0067$), since that is relatively free from the back-reactions of renormalized energy-momentum tensors.

\begin{figure}
\begin{center}
\includegraphics[scale=0.9]{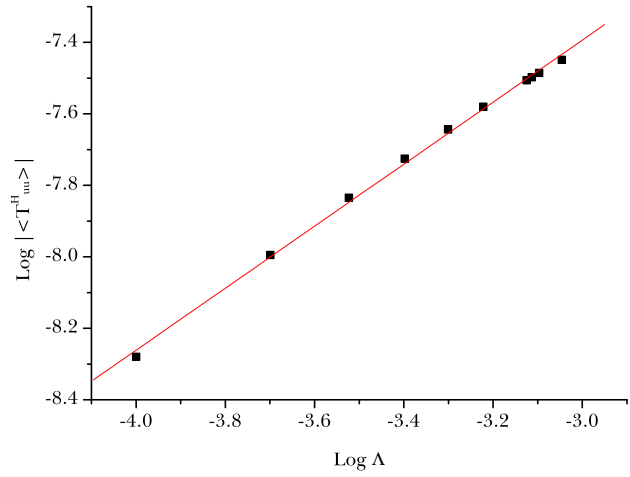}
\caption{\label{fig:test}A log-log plot of $|\langle\hat{T}^{\mathrm{H}}_{uu}\rangle|$ and $\Lambda$. The gradient is $0.86789 \pm 0.01333$ and approximately $1$.}
\end{center}
\end{figure}

Figure~\ref{fig:test} is a plot for the maximum of $|\langle\hat{T}^{\mathrm{H}}_{uu}\rangle|$ along an asymptotic surface ($u \simeq 0.0067$) as varying $\Lambda = 0.0001, 0.0002, 0.0003, 0.0004, 0.0005, 0.0006, 0.00075, 0.00077, 0.0008, 0.0009$. The gradient is $0.86789 \pm 0.01333$ and approximately $1$. Note that the smallest value of $\langle\hat{T}^{\mathrm{H}}_{uu}\rangle$ is observed on an almost same radius. We analytically expect that the renormalized energy-momentum tensors in a de Sitter background should be \cite{Birrell:1982ix}
\begin{eqnarray}
\langle\hat{T}^{\mathrm{H}}_{uu}\rangle \propto -P \frac{\Lambda}{r^{2}_{0}},
\end{eqnarray}
where $r_{0}$ is a certain constant radius. Therefore, we expect that the proportionality will be approximately $1$. This correspondence is not so trivial since our numerical setup does not contain a direct relation between $\Lambda$ and $\langle\hat{T}^{\mathrm{H}}_{uu}\rangle$. Therefore, we can conclude that our numerical calculations exhibit this behavior consistently, even though there are some ambiguities.

\end{document}